# Numerical Study of the Mixing Process of Unconfined Diffusion Flame in a Laminar to Transition-to-Turbulent Regime in a Four-Port Array Burner

M. De la Cruz-Ávila [a*], J. E. De León-Ruiz [b], E. Martínez-Espinosa [a], I. Carvajal-Mariscal [b], and L. Di G. Sigalotti [c]

[a] Universidad Nacional Autónoma de México, Instituto de Ingeniería, Ciudad Universitaria, 04510 México City, MÉXICO.
[b] Instituto Politécnico Nacional, ESIME UPALM, Av. IPN s/n, 07738 México City, MÉXICO
[c] Universidad Autónoma Metropolitana-Azcapotzalco, Av. San Pablo 180, 02200 México City, MÉXICO
*Correspondence: mauriciodlca1@gmail.com

**ABSTRACT**

The mixing process of multiple jets of liquefied petroleum gas and air in a diffusion flame is numerically analysed. The case study considers a four-port array burner where the fuel is injected by four peripheral nozzles and mixed with the surrounding air. Simulations are conducted with the Reynolds-Averaged Navier-Stokes technique, and the turbulence effect is modelled with the realizable k-e model. In addition, the eddy dissipation model is implemented to calculate the effect of the turbulent chemical reaction rate. Results show that the essential mixture mechanism occurs within a flame cone derived from the fuel jets interaction. However, the mixing process is driven by jets' drag allowing an air/fuel smooth mixture to reach the flammability limits at two zones: one at a lower location or close to the burner surface and a second before the flame front development. The entire mixing mechanism culminates with the development of the flame necking cone. Any air concentration that falls into the cone radius will be entrained, contributing to the overall flame structure. Since the cone radius reach is limited only by radial distance of the jet array and the nozzles' distance, the flame heights, consequently, depend solely on fuel mass flow.

**Keywords:** *Multiple jet mixing process, Port array burner, Unconfined diffusion flame, LPG Combustion, Numerical simulations.*

## 1 INTRODUCTION

Understanding that fossil fuels are still the main energy source around the world, the need to achieve a more efficient performance of the technologies developed for power generation is imperative, thus leading to a cleaner output decreasing its impact on the environment. Based on this, the utilization of diffusion flames has become more relevant since, given their nature, they have a wide range of combustion even under turbulent conditions than their premixed or co-flow counterpart. However, the high soot [1,2], as well as the $NO_x$ and $SO_x$ load characteristics, limit the implementation of these flames for both industrial and domestic applications.

Yamamoto et al. [3] investigated diffusion flames in a triple port burner pointing that, by increasing the external air flow velocity, inner flame re-attachment and flip-flop between inner and outer flames behaviours were observed, and the most important, when the flame is lifted, the maximum soot concentration and $NO_x$ emission are decreased. Because of that argument, the construction of industrial and commercial combustors searches to make a short flame at high heating loads. One solution would be the use of multiple nozzles, which reduces the fuel flow rate per nozzle because laminar diffusion flame length is linearly proportional to the fuel flow rate according to Roper [4]. Delichatsios [5] argued that using a small-diameter nozzle is one method to get reduced turbulent flame length because it is proportional to the nozzle diameter and not to the fuel flow rate. However, Kalghatgi [6]explained that a small-diameter nozzle cannot burn as much fuel as a large-diameter one because the turbulent blowout velocity, $u_{bl}$, is proportional to the jet diameter. Therefore, to afford the high heating load requirement and small flame length, using multiple nozzles would be a solution.

The multiple-jet array [7] has demonstrated many advantages over single/co-flow diffusion flames. When a compact combustor with multiple nozzles is used, it is unavoidable the flames' interaction with each other. This interaction is altered by the geometrical arrangements, the number of fuel jets, and the space between nozzles, which change the merging height, blowout velocity, concentration profiles, among other particularities related to the overall flame development. To mention, Raghunathan and Reid [8]argued that stream mixing is improved and large noise reduction is achieved. Numerous studies based on the hydrodynamic development of non-reactive multi-jets have been reported in the literature [9–14]. However, little research has been done on reactive flows for multi-jet configurations. Menon and Gollahalli [15,16], studied the interaction of 2, 3 and 5 propane jet flames. Leyte et al. [17] were interested in the effect of the number of jets, spacing between jets, and the tube diameter on the acetylene flame length. Lenze et al. [18] who were able to correlate the ratio of the attached flame length of multi-jets to a single jet with the separation distance and the number of jets, studied the mutual effects of 3 and 5 jet diffusion flames on a city gas and natural gas burner. Those measurements are related to multiple free and trapped flame concentrations, flame length and width. Nonetheless, the mixing mechanism for reacting flows has been not sufficiently explained. In a previously published research [19], the overall subject was fragmented into several parts and a thorough analysis of each was presented to improve the understanding of the involved phenomena. A numerical study of a confined flame was conducted, based on the output visualization and data, the resulting flame was then divided into different layers and zones, where their importance and role are comprehensibly stated and analysed.

In this way, important investigations show the analysis of diffusive flame structures, especially, presenting the effects of global hydrodynamics during the flame front development [20]. However, they scarcely show how these physical mechanisms arise. In theses, the heat release, the flame power and the species concentrations at certain positions are analysed. However, when placed on the physical analysis, they present many assumptions and conjectures of the mixing process.

Seepana and Jayanti [21] show an exhaustive analysis of the flame structure in fuels, mainly using the chemical kinetics approach based on the concentrations of the most important and stable species. Another example is the work of Masri et al. [22] that shows the structure of turbulent non premixed flame using the Rayleigh/LIF visualization technique and whose work also focuses on the analysis of the turbulence effects specifically, over the mixture fraction without showing where this turbulence comes from or what could cause it. Still others quantify the internal and external vortices interaction [23] and how they help in the mixing process, but scarcely show how or under what conditions these mixing zones occur. The information from these works, like many others, is extremely valuable, however little is said about the physical causes that intervene in the mixing process and less, how this mixing process occurs in the lower area of the burners.

It could be assumed that chemical reactions are the most interesting or helpful to analyse. Although this statement is useful, it is also important to analyse how it is that, given these chemical reactions continue to develop, they help to preserve the flame front, but analysed from another point of view. If this analysis of the mixing process is also approached from the physical perspective, it is possible to better understand how the interactions of the three main parameters are presented in the development of diffusion flames such as chemical kinetics, global hydrodynamic and thermal mechanisms, which are interrelated and impossible to separate. Decoupling these mechanisms would cause a bias of valuable information. It is a misconception to decouple and analyse the reactive flows as if they were non-reactive flows since chemical kinetics and thermal mechanisms participate in the development, since they interact with each other whenever there are these types of flames.

For all these reasons, the current numerical study aims on the mixing process of an unconfined Liquefied Petroleum Gas diffusion flame in a 4-port array nozzle distribution in which the interaction of the fuel injection jets is analysed and shows the monitoring of such physical effects, without decoupling the three main parameters that constitute the flame development. The proposed analysis allows the visualization and identification of the relevant areas that interact between each nozzle and favours the combustion process taking account the chemical, hydrodynamic and thermodynamic mechanisms. The numerical approach, of the three aforementioned parameters, is presented in a tenuous way due to the great demand of computational resources that this type of study implies. The study is conducted employing six different fluid flows ranging from laminar until transition-to-turbulent flame regimes. This study can help avoid several inverse diffusion flame combustion problems, including uncontrolled flame lifting, flickering, undesired flame stretching, local extinction, and flame blowout, commonly present in industrial applications, such as direct-contact steam generation for in situ combustion helping to improve the efficiency of these technologies.

## 2  EXPERIMENTAL SETUP

The experimental-bench uses a 4-port-nozzle gas-burner injection system to study the behaviour and define the structure of a laminar to transition-to-turbulent unconfined diffusion flame, which was built based on a Lug-Bolt distribution design [19] as shown in Fig. 1a. It consist of four 0.8mm peripheral nozzles in a radial distribution configuration $Rd$=4x16.94mm, a diameter gas-burner $d$=25.4mm, with a nozzle to nozzle distance $S$=11.98mm as seen in Fig. 1b.

To avoid a drastic pressure drop, the cross-sections of the inlet gas tubes were determined to relate length and diameter near the most stoichiometric inlet velocity values. The nozzles diameters were calculated to maintain the ratio between air and fuel closest to the stoichiometric, which are summarized in Table 1. The entire experimental rig was designed to reduce both, the thermal stress of the combustion phenomenon and pressure-drop in the fuel side employing symmetric dimensions and proportional magnitudes. Due to the scope of this evaluation, the 4mm central-nozzle was not used and left aside as a complement of an ongoing research project involving a complete Lug-Bolt configuration [19].

The experimental assessment was not only designed to determine the flame characteristics but, also its hydrodynamic behaviour utilizing six different fuel flows: (A) 350cc/min, (B) 650cc/min, (C) 950cc/min, (D) 1200cc/min, (E) 1500cc/min and (F) 1800cc/min as seen in Table 1.

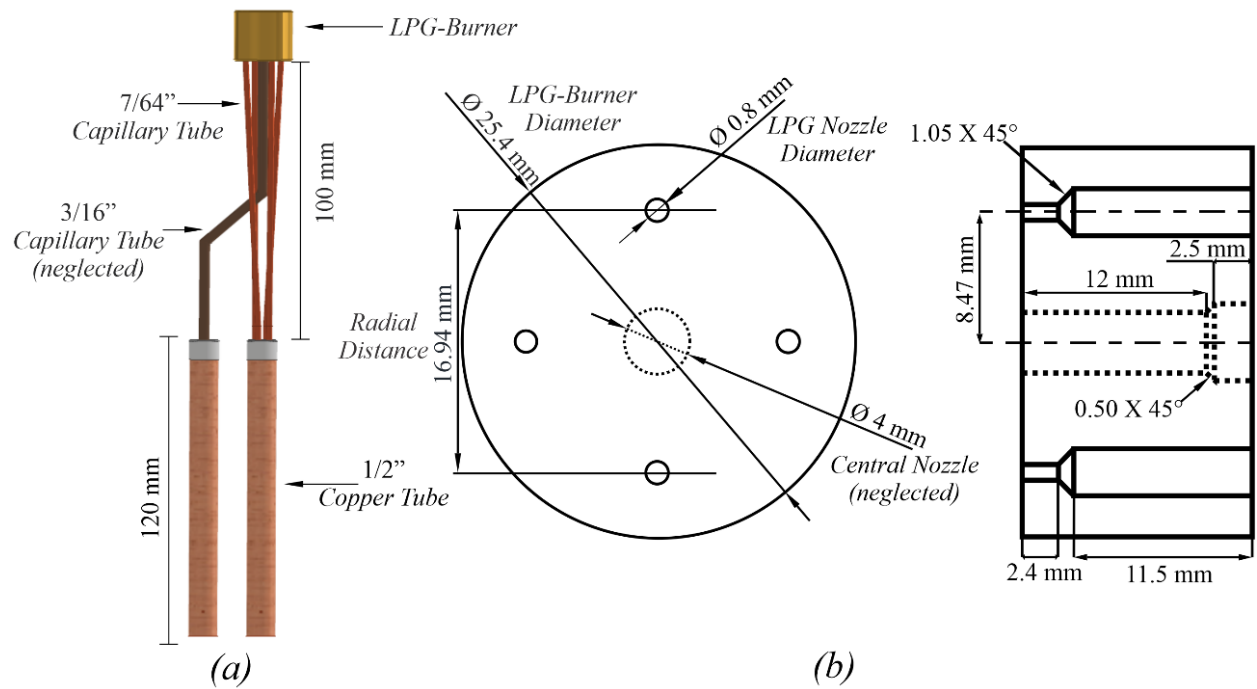

Fig. 1: Central-peripheral fuel injection system: (a) distribution setup, (b) gas-burner configuration.

For this experimental study, liquefied petroleum gas (LPG) was used containing approximately 60% propane and 40% butane; composition with thermochemical properties similar to those reported by Mishra and Rahman [24]. The surrounding air was employed as the oxidizer for the unconfined diffusion flame.

In the well know combustion theory for a hydrocarbon fuel represented by $C_xH_y$, the stoichiometric relation is,

$$C_xH_y + a(O_2 + 3.76N_2) \rightarrow xCO_2 + \left(\frac{y}{2}\right)H_2O + 3.76aN_2. \qquad (1)$$

For simplicity, in the above reaction the air has a composition of 21%$O_2$ and 79%$N_2$. It is assumed that this reaction is balanced, where a=x+(y/4), then the stoichiometric air/fuel ratio would be $(A/F)_{stoich}=(m_{ox}/m_{fuel})=4.76a(MW_{air}/MW_{fuel})$ where $MW_{air}$ and $MW_{fuel}$ are the molecular weights of air and fuel, respectively.

Table 1: LPG-air hydrodynamics properties at 293.15K and 0.7647atm.

| Case Studies | A | B | C | D | E | F |
|---|---|---|---|---|---|---|
| LPG Volumetric Flow [cc/min] | 350.0 | 650.0 | 950.0 | 1200.0 | 1500.0 | 1800.0 |
| LPG Injection Velocity [m/s], $V_c$ | 2.902 | 5.389 | 7.875 | 9.948 | 12.434 | 14.921 |
| LPG Mass Flow [kg/s] | 8.91x10$^{-6}$ | 1.65x10$^{-5}$ | 2.41x10$^{-5}$ | 3.05x10$^{-5}$ | 3.81x10$^{-5}$ | 4.58x10$^{-5}$ |
| LPG Injection Re | 498.1 | 925 | 1351.6 | 1707.4 | 2134 | 2561 |
| Stoichiometric Air Mass Flow needed [kg/s] | 1.67x10$^{-4}$ | 3.10x10$^{-4}$ | 4.54x10$^{-4}$ | 5.73x10$^{-4}$ | 7.17x10$^{-4}$ | 8.61x10$^{-4}$ |
| Mass Air–Fuel Ratio $(A/F)_{stoich}$ | 15.5 | | | | | |

To measure the flame temperature values, a Fluke-Ti55FT thermal-imaging camera was located in the workbench frontal-plane at a 1m from the gas-burner centre avoiding emissivity errors [25–27] where several images were taken and analysed. This camera provides quantitative information about the flame's maximum temperature zone. Furthermore, the vertical temperature profile was measured along the flame axis, from the base, up to the unstable flame tip, employing a Heraeus pyrometer Mod. DT-400 (1% F.S.) with Tungsten-Rhenium alloy thermocouple probe positioned along the axial centreline above the nozzle.

These particular alloy type C thermocouples are mostly used for measuring temperatures from 293K up to 3073K due to their properties and high melting point around 3373K. Additionally, have demonstrated reliability, measurement consistency and durability for temperature ranges of 1473 to 1723K with 1000h [28], 1673 to 2173K with several hundred hours [29] and 2593K of 240h [30].

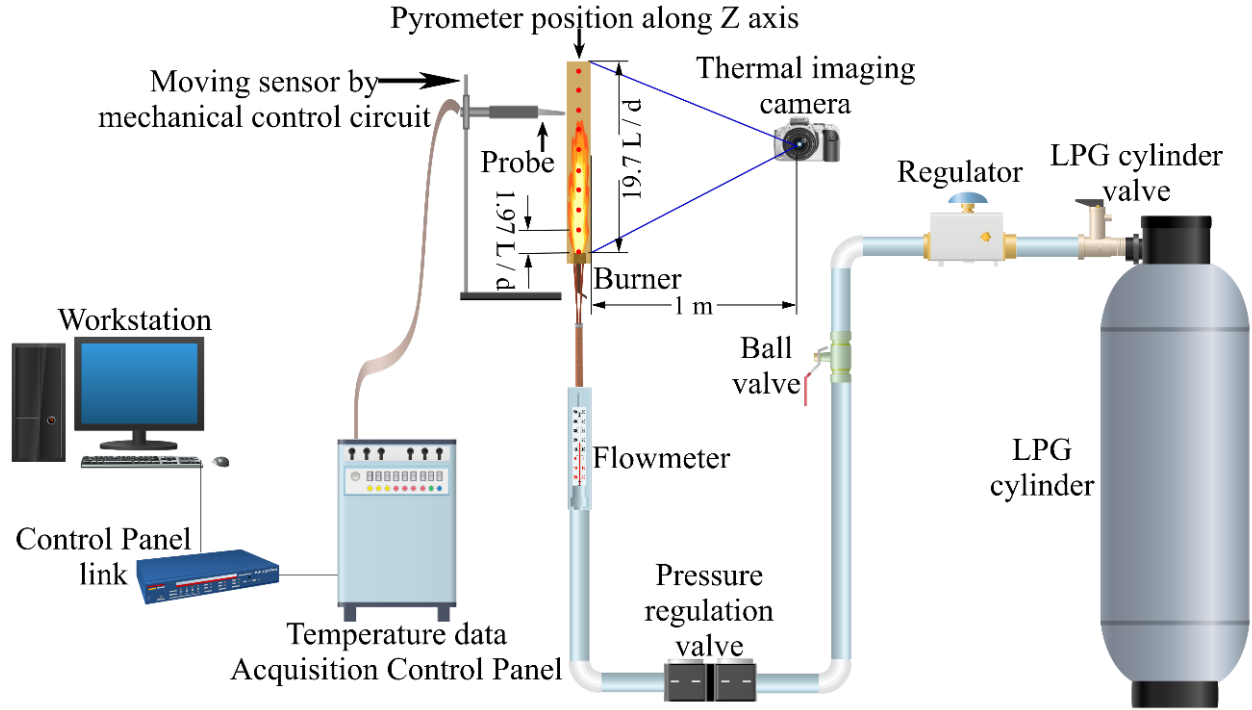

Fig 2: Diagram of the validation experimental setup.

The temperature was determined through 6s measurements at ten different positions along with the developed flame, as shown in Fig. 2. In order to maintain a measurement error below ~10% and to ensure that said measurement lies within a 95% confidence interval, CI, it was determined that a minimum of 288 samples was required, which given the experimental design factors and levels, entailed at least 5 replications. However, given the rather straightforward setup, the value was four times higher, amounting to 20 samples per position and this number of treatments for every flow condition was adopted.

## 3 NUMERICAL SIMULATIONS
### 3.1 CASE STUDY

The numerical model consider an unconfined LPG-air diffusion flame employing the dimensions of the burner configuration described in Fig. 1. The nozzles are established in a radial distribution to maintain air entrainment to the stoichiometric relation. The virtual combustion domain has a diameter $D$=150 mm and total length $L$=500 mm. The injection nozzles have the exact geometrical array of 4 x 16.94 mm radial distribution with exit nozzles diameters $d_f$ =0.8 mm as shown in Fig. 1a. The numerical study matrix is performed for the six different fluid flows with thermodynamic conditions listed in Table 2. The analysis is focused in two precise plane zones in which data was obtained by mean of *Markers*. The zones are intended to cover the XZ central-plane and 45º rotation over the Z-axis to place another XZ angled-plane in which mixing layers contours are significant and 10 *Markers* with a 50mm separation distance between them illustrated in Fig. 3a to capture all properties needed for streams mixing and flame development, according to experimental set up of Fig. 2.

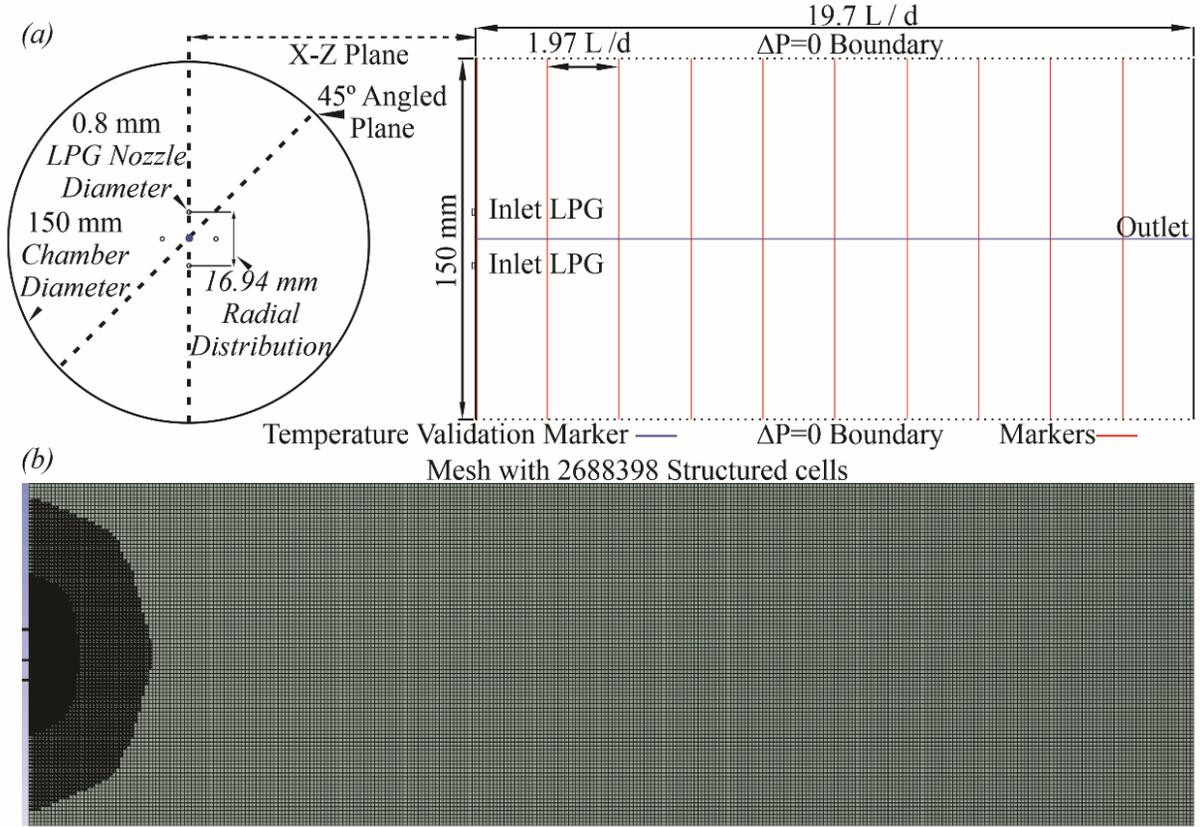

Fig. 3: (a) Geometry details for numerical simulations; (b) mesh details.

## 3.2 NUMERICAL DETAILS

The numerical simulations consider a combination of the advancing-front meshing [31,32] and structured cell methods. One of the advantages offered by the advancing-front method over commonly structured grids is the facilitating *tessellation* process for geometrically complicated domains allowing the mesh density to be adapted to the geometry. These merging methods result in a growth of thin layers and cells from the downward wall of the combustion chamber to the final domain edge, allowing the implementation of high-order discretization schemes.

Besides these meshing methods, an adaptive time-stepping was considered to ensure the correct develop-time in complex simulations as occurs in combustion process, where high speeds and sudden energy release are present. Table 3 shows the time-step computed through equation (2), which must be small enough to solve time-dependent features and to ensure convergence within several iterations.

$$\Delta t \approx \frac{Typical\ cell\ size}{Characteristic\ flow\ velocity} \qquad (2)$$

Table 3: Time stepping characteristics for temporal discretization on combustion process.

| º | A | B | C | D | E | F |
|---|---|---|---|---|---|---|
| Fixed | | | | $1 \times 10^{-4}$ | | |
| Adaptive | $1 \times 10^{-2} > \Delta t > 1 \times 10^{-4}$ | $7 \times 10^{-3} > \Delta t > 1 \times 10^{-4}$ | $4 \times 10^{-3} > \Delta t > 1 \times 10^{-4}$ | $1 \times 10^{-3} > \Delta t > 1 \times 10^{-4}$ | $7 \times 10^{-4} > \Delta t > 1 \times 10^{-5}$ | $4 \times 10^{-4} > \Delta t > 1 \times 10^{-5}$ |

The simulations considered a mass flow injection of each case (see Table 1). A zero-pressure gradient condition is implemented at the outlet boundary and the gas discharges are allowed to occur at reduced atmospheric conditions of 0.7647atm and temperature of 293K. Furthermore, the Monotonic Upwind Scheme for Conservation Laws (MUSCL) scheme [33] was used for evaluating the convective and viscous terms, which provides a highly accurate numerical solution for the system. For pressure-velocity coupling, the Coupled Solution Method algorithm was employed, which solves the governing equations of continuity, momentum, energy and species transport simultaneously. Since the momentum and continuity equations are solved in a closely coupled manner, the rate of solution convergence significantly improves when compared to the segregated algorithm. However, memory usage increases from 1.5 to 2 times, compared against the segregated algorithm PISO, since the pressure-based continuity equation system must be stored in memory by solving velocity and pressure fields instead of a simple equation, as is the case with the segregation algorithm. All the numerical simulations were performed with an academic license Ansys Fluent v13 software.

The Fig. 3b shows the resulting mesh with 2688398 structured cells, is between the range where variations by numerical diffusion due to cell size does not represent a significant improvement to the solution, but it does present a visual improvement of the results obtained. The sensitivity analysis details of the described geometry and the combustion phenomena involved are further described in a previous work [34].

### 3.3 GOVERNING EQUATIONS

The equations to be solved are the conservation laws of mass, momentum, energy and chemical species. A density-weighted averaging or Favre-averaging denoted by "~" is considered because the variable density attached to the combustion phenomena models gives more accurate predictions used along with the time average designed by "¯". These terms are used in the transport equations to model the flow movement and the heat exchange. The Favre-averaged continuity, momentum, energy and species equations are expressed as follows:

*Mass conservation:*

$$\frac{\partial \bar{\rho}}{\partial t} + \nabla \cdot (\bar{\rho}\tilde{\mathbf{u}}) = 0, \tag{3}$$

*Momentum:*

$$\bar{\rho}\frac{\partial \tilde{\mathbf{u}}}{\partial t} + \bar{\rho}\tilde{\mathbf{u}} \cdot \nabla \tilde{\mathbf{u}} = -(\nabla \bar{p}) + \nabla \cdot \bar{\tau} + \bar{\rho}\sum_{i=1}^{N} \widetilde{Y_i \mathbf{f}_i} - \nabla \cdot (\bar{\rho}\widetilde{\mathbf{u}'\mathbf{u}'}), \tag{4}$$

*Conservation of Energy:*

$$\bar{\rho}\frac{\partial \tilde{e}}{\partial t} + \bar{\rho}\tilde{\mathbf{u}} \cdot \nabla \tilde{e} = -\nabla \cdot \tilde{\mathbf{q}} - \overline{p\nabla \cdot \mathbf{u}} + \overline{\tau:\nabla\mathbf{u}} + \bar{\rho}\sum_{i=1}^{N} \widetilde{Y_i \mathbf{f}_i \cdot \mathbf{V}_i} - \nabla \cdot (\bar{\rho}\widetilde{\mathbf{u}'e'}), \tag{5}$$

*Species:*

$$\bar{\rho}\frac{\partial \tilde{Y}_i}{\partial t} + \bar{\rho}\tilde{\mathbf{u}} \cdot \nabla \tilde{Y}_i = \nabla \cdot \left(-\bar{\rho}\widetilde{\mathbf{V}_i Y_i}\right) + \bar{\omega}_i - \nabla \cdot (\bar{\rho}\widetilde{\mathbf{u}'Y_i'}), \quad i = 1, \dots, N, \tag{6}$$

where $\bar{\rho}\widetilde{\mathbf{u}'\mathbf{u}'}$, $\bar{\rho}\widetilde{\mathbf{u}'Y_i'}$ and $\bar{\rho}\widetilde{\mathbf{u}'e'}$ are the Reynolds stress tensor, mass-weight density fluctuations, and turbulent heat transfer vector, respectively. $\omega_i$ is the $i^{th}$ species production rate and $e$ may be expressed as $e = \sum_{i=1}^{N} h_i Y_i - p/\rho$. The term $\widetilde{\mathbf{u}'Y_i'} = \overline{\rho\mathbf{u}'Y_i'}/\bar{\rho}$ is the Favre mean Reynolds flux for $Y_i$ representing the process of large-scale mixing by turbulent transport and generally modelled according to a gradient transport approximation

$$\widetilde{\mathbf{u}'Y_i'} = -\frac{\mu_T}{\bar{\rho}\sigma_i}\nabla\widetilde{Y}_i, \tag{7}$$

where $\mu_T$ is the eddy viscosity and $\sigma_i$ is the turbulent Schmidt number. It is commonly use $\mu_T = c_\mu \bar{\rho}\widetilde{k}^2/\widetilde{\varepsilon}$ where $c_\mu$ is managed by the turbulence model as constant in standard $k$-$\varepsilon$ turbulence model or variable in the Realizable $k$-$\varepsilon$ turbulence model. $\widetilde{k}$ and $\widetilde{\varepsilon}$ are the mean turbulence kinetic energy and dissipation rate, respectively. For numerical solution these equations are written in Cartesian coordinates.

Equations (4), (5), and (6) demand additional mathematical expressions for terms represented by the general form $\bar{\rho}\widetilde{\boldsymbol{\phi}'\mathbf{u}'}$. The equations closure requires modelling of the Reynolds stress tensor, turbulent heat flux, and mass-weight density fluctuations. Then, the Reynolds stress tensor is closed with turbulence model Realizable $k$-$\varepsilon$ [35]. The turbulent heat flux vector and mass-weight density fluctuations are obtained through an analogy between momentum transfer and molecular diffusion. The Realizable $k$-$\varepsilon$ turbulence model [35,36] has been validated experimentally for many reactive flows with satisfactory results [23,37,38]. This model is analogous to the standard $k$-$\varepsilon$ model but $C_\mu$ is managed as variable, which represents an important prediction approach for the viscous terms and more precision in combustion analysis.

## 3.4 COMBUSTION MODELING DETAILS

The species are introduced by means of their mass fractions Yi for i=1 to N, where N specifies their number in the reactive mixture. The mass fractions, Yi, are defined by Eq. (8), where $m_i$ is the mass of species $i$ present in a given volume $V$ and $m$ is the total mass of gas in the volume

$$Y_i = \frac{m_i}{m}. \tag{8}$$

The energy production of LPG is established by the overall reaction given by Eq. (9), which implies a rather robust simplification of the actual reaction mechanism that involves many free-radical chain reactions. Nevertheless, the main purpose presented in this research is not to analyse the secondary chemical reactions. For this reason, a single-step irreversible chemical reaction was used to focus on the flow development,

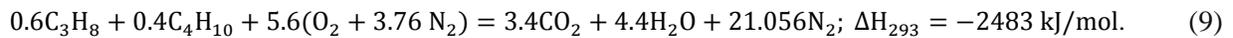

$$0.6\,C_3H_8 + 0.4\,C_4H_{10} + 5.6(O_2 + 3.76\,N_2) = 3.4\,CO_2 + 4.4\,H_2O + 21.056\,N_2;\ \Delta H_{293} = -2483\ \text{kJ/mol}. \tag{9}$$

The turbulent chemical reaction rate is modelled with the Eddy Dissipation Model (EDM) [39], which is based on the infinitely fast chemistry hypothesis and assumes that the reaction rate is controlled by the turbulent mixing. The EDM generalized formulation has been proposed to take into account finite-rate chemistry effects. A stoichiometric relation describing chemical reactions of arbitrary complexity can be represented by the $r^{th}$ reaction equation [40]. The species $i$ production net-rate, $R_{i,r}$, due to reaction $r$, is given by the smaller, *limiting-value*, of the two expressions below, which are based on reactants and products mass fraction given by Eqs. (10) and (11), where $Y_p$ and $Y_R$ the species mass fraction of products and reactants respectively, $A$ and $B$ are Magnussen [39] constant for reactants (4.0) and products (0.5) respectively. $M_{w,i}$ is the molecular weight for both $R$ and $P$ for reactants and products respectively.

$$R_{i,r} = v'_{i,r} M_{w,i} A \rho \frac{\varepsilon}{k} \min_{\mathcal{R}} \left( \frac{Y_{\mathcal{R}}}{v'_{\mathcal{R},r} M_{w,\mathcal{R}}} \right), \quad (10)$$

$$R_{i,r} = v'_{i,r} M_{w,i} A B \rho \frac{\varepsilon}{k} \frac{\sum_P Y_P}{\sum_j^N v''_{j,r} M_{w,j}}. \quad (11)$$

### 3.5  COMBUSTION MODELING VALIDATION

If the combustion reaction is carried out under adiabatic conditions then $\delta Q=0$ at constant pressure, the first law of thermodynamics yields $dH=0$, where the energy released in the combustion process raises the thermal level of the reaction products. Therefore, the burnt and unburnt gasses, index $b$ and $u$ respectively, have the same specific enthalpy where it is verified that:

$$\delta Q = 0 \rightarrow dH = 0 \rightarrow H^b = H^u, \quad (12)$$

where $Q$ is the heat, $H$ and $h$ are the enthalpies. Furthermore, the molar enthalpies of the burnt and unburnt gasses often differ, because the amount of molecules usually changes in a chemical reaction. Thus,

$$h^u = \sum_{j=1}^{S} w_j^u h_j^u = \sum_{j=1}^{S} w_j^b h_j^b = h^b, \quad (13)$$

where $h$ are the enthalpies $j$ denotes the species and $w$ the work. For constant pressure the relation holds:

$$h_j^b = h_j^u + \int_{T_u}^{T_b} c_{p,j} dT. \quad (14)$$

The adiabatic flame temperature, $T_{ad}$, or maximum flame temperature, is the one reached when the combustion process in adiabatic conditions corresponds to a complete-combustion of a stoichiometric mixture, which can be obtained from the balance established between the enthalpy of both burnt and unburnt gasses. Under any other conditions, the flame temperature, $T_f$, will vary, is by this that in particular experiments conditions $T_f$ will have always a lower value. In order to evaluate the $T_{ad}$ it is necessary to know the composition of both the reactive and products' mixtures, involved this process. Using the equation (14) the $T_{ad}$ can be determined, i.e., the temperature resulting after combustion provided that heat losses to the surroundings are negligible. The LPG adiabatic flame temperature, $LPG\text{-}T_{ad}$, can be computed by a simple iteration method. Because both temperatures describe a similar behaviour with the difference that they are carried out under different conditions, $T_f$ and $T_{ad}$ maintain a relation by which the global performance of the flame is described, and for this reason, $T_f/T_{ad}$ is used as a practical way to parameterize that relation.

In commonly used combustion configurations, even with the well-controlled experimental area and guaranteed mixing process they are exposed to minimal air variations. This causes perceptible modifications that lowers $T_f$ and the corresponding measurement compared to that obtained through analytical modelling produces a lower output. For this particular study, the resulting theoretical magnitude is $T_{ad}=2398K$, which is consistent within the range reported by Silverman [25]. The average discrepancy between the experimental and numerical datasets is below to an overall 6.6%, which considered relatively not significant, based on the order of temperature magnitude reached by the flame. Table 4 shows the $T_f$, comparison and variation at distinct positions between experimental and numerical dataset.

Table 4: Maximum $T_f$ measured in K.

| Case | A | B | C | D | E | F |
|---|---|---|---|---|---|---|
| Fuel flow [cc/min] | 350 | 650 | 950 | 1200 | 1500 | 1800 |
| Thermal imaging | 1955 | 2202 | 2115 | 2153 | 2214 | 2131 |
| Numerical Simulations | 1897 | 2157 | 2072 | 2164 | 2148 | 2088 |
| Pyrometer | 1944 | 2029 | 2109 | 2231 | 2200 | 2126 |
| Position L/d | 1.97 | 5.9 | 5.9 | 9.85 | 9.85 | 5.9 |

In the context of a fuel mixture, the auto-ignition temperature, $T_{ai}$, is the lowest temperature at which the fuel will ignite spontaneously in air at atmospheric pressure without the aid external ignition energy source, such as a flame or spark. It gives a temperature indication at which a material will spontaneously burst into flames when exposed to the atmosphere. Raising the fuel temperature to its self-ignition point provides the energy necessary to initiate the chemical reaction for combustion. In all cases and given fuel mixture composition described in Section 2, the $T_{ai}$=775K or $T_{ai}$=0.2$T_f$/$T_{ad}$.

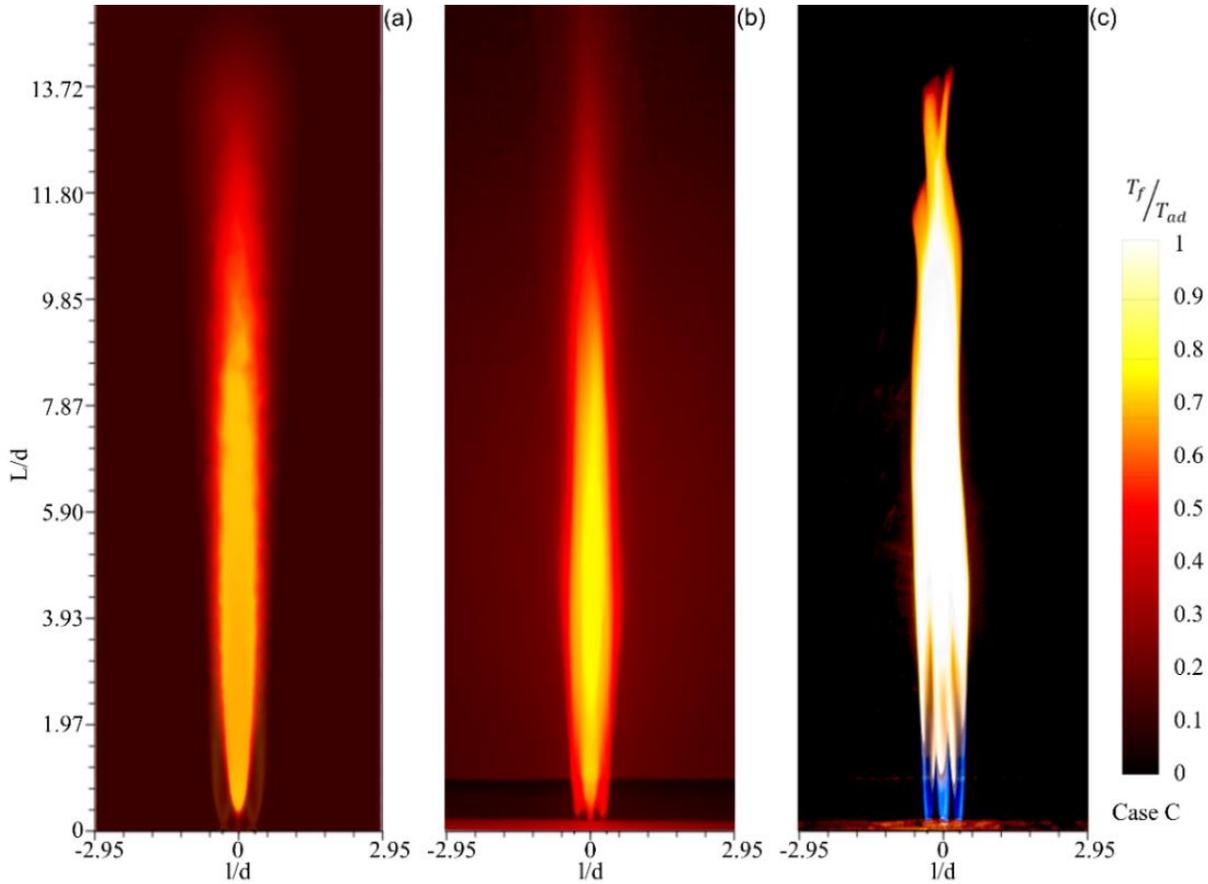

Fig. 4: Flame front border temperature profiles.
a) Numerical simulation; b) Thermal-imaging; c) Experimental instantaneous image.

All the numerical predictions were compared against experimental temperature measurements. In Fig. 4 it is possible to observe the comparison between the three instantaneous images of the temperature profile along the Z-axis of Case C. The selection of this particular Case has been since, during the acquisition of the temperature measurements, the data collected by the pyrometer in each different position there was substantial temperature variation, that is, in each position, there was measurement temperature including the furthest located at 17.71L/d higher than the $T_{ai}$. This means that it is either a temperature from the flame front or a combustion products' temperature, in which case it was also reported. In both cases, Case A and B respectively, the pyrometer obtained a temperature reading less than the $T_{ai}$ indicating that it is a temperature directly from the combustion products. For Cases D, E, and F, the temperature measurement at the last data acquisition position 17.71L/d were: $0.29T_f/T_{ad}$, $0.42T_f/T_{ad}$ and $0.52T_f/T_{ad}$ respectively. Because each of these values exceeds the $T_{ai}$, it is known that they are direct flame front temperature readings. Therefore, case C was the most suitable because, in all the positions to which the pyrometer was placed, it completely covered not only the flame front but the temperature of the flame tip.

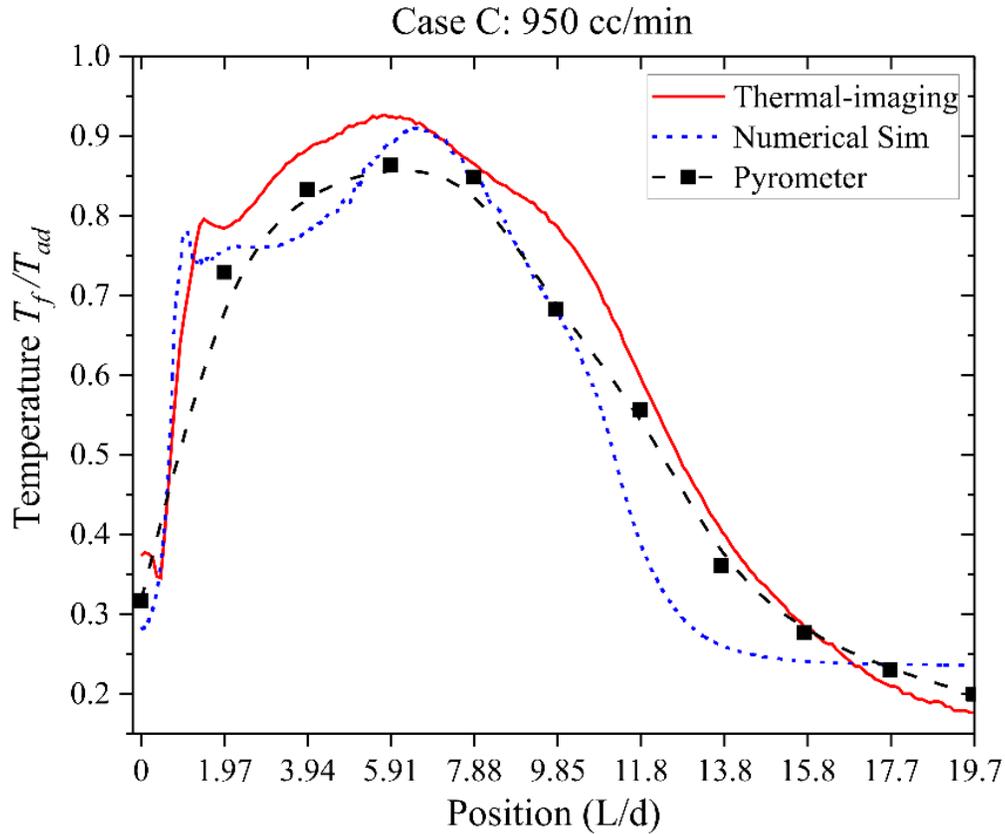

Fig. 5: Flame front temperature profiles.
a) Numerical simulation, b) Thermal-imaging, c) Experimental through pyrometer acquisition.

Fig. 5 presents the temperature profile along the Z-axis comparing the three datasets. The continuous red line represents the flame front temperature extracted directly from the thermal-imaging. The dots blue line exhibits the numerical prediction of the flame front temperature. The dashed line shows the curve tendency of the mean average values of the temperature measured by the pyrometer.

A first data that is perceived from the temperature profile development is in the range of 0 to 0.98L/d with a value of $0.376T_f/T_{ad}$, captured only by the thermography. However, at the 0L/d position in which the first location the pyrometer was placed, it could not extract the information. Also, this temperature variation was not obtained through numerical simulations due to the equations averaging on which the RANS technique is based. Therefore, the thermal imaging data is used as the higher temperatures profile benchmark and pyrometer as the overall behaviour to which the numerical result must approach.

Besides, this figure shows, in the range of 0<1.97L/d>3.94 a sudden temperature increment peak. This behaviour is captured by the thermal imaging and well simulated by numerical predictions. However, due to the pyrometer position, this information could not be collected. After this point, the temperature, for the three data groups, grows to its maximum value in the range of 3.94L/d to 7.88L/d. Then, the temperature cools down indicating that combustion, i.e. reaction is over and the combustion products are dispersing off. The uncertainty measurement for the pyrometer are presented in table 5.

Table 5: Uncertainty Pyrometer measurement results

| Source | Position | Replications | Mean | Std Dev | Statistical tolerance | Expanded Uncertainty | Relative Uncertainty |
|---|---|---|---|---|---|---|---|
| [cc/min] | [L/d] | | [K] | [K] | [K] | [K] | % |
| | 0 | 20 | 760.15 | 7.07 | | | |
| | 1.97 | 20 | 1750.05 | 5.52 | | | |
| | 3.93 | 20 | 1993.9 | 11.73 | | | |
| | 5.9 | 20 | 2070.7 | 14.19 | | | |
| | 7.87 | 20 | 2033.1 | 8.85 | | | |
| 950 | 9.81 | 20 | 1637.2 | 8.54 | 7.05 | 4.25 | 1.64 |
| | 11.8 | 20 | 1334.65 | 9.65 | | | |
| | 13.72 | 20 | 864.6 | 6.89 | | | |
| | 15.68 | 20 | 663.7 | 5.42 | | | |
| | 17.71 | 20 | 550.5 | 13.31 | | | |
| | 19.67 | 20 | 481.85 | 8.79 | | | |

By comparing the datasets obtained, the difference between the numerical results and the experimental measurements has an overall temperature deviation of 7.4% and 5.7% concerning the position. Therefore, the numerical simulation results are suitable for the mixing process analysis since they agree with the most complex mechanism, that is to say, the combustion representation process agreeing with the experimental measurements. It is worth mentioning that, these deviations could be reduced by using a combustion model that considers all the chain reactions, but the calculation time could be 20 or 30 times greater.

## 4 RESULTS AND DISCUSSION
### 4.1 REACHING THE FLAMABILITY LIMITS

In combustion phenomena, especially in diffusion flames, the mixing process is highly relevant because the air/fuel ratio must remain stable, and on the other hand, it must be ensured that the stoichiometric mixture is controlled only by the fuel mass flow.

As a first step to estimate the mixing process, is to identify a special characteristic of the flows named potential core, then analyse how jets develop after this region. The potential core is defined as the region in the jet in which the centreline velocity remains essentially constant and equal to the centreline velocity at the nozzle exit. The main importance of this feature is that any detriment of the general behaviour on the development and mixing process is located just after the reach and evolution of the potential core depicted in Fig 6. This feature is equal to the inlet velocity of each case illustrated in red over the dimensionless velocity contours calculated as the case velocity divide by the maximum case velocity $V_c/V_{cmax}$. These contours are extracted by means of a cross-section on the central-plane along the Z-axis. After identifying this region, the next step is to decompose the flame jet in zones as explained below.

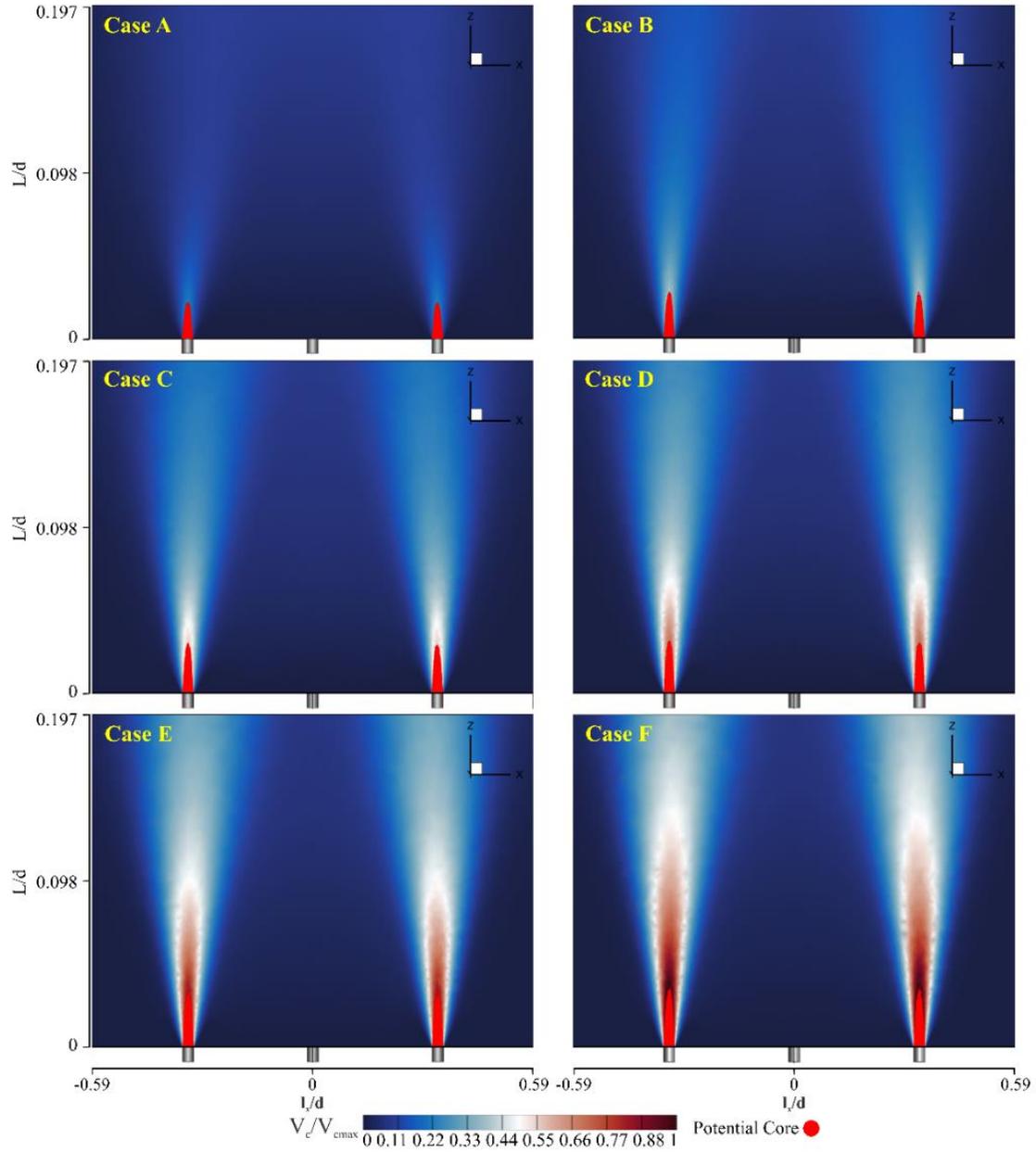

Fig. 6: Potential core close-up for each case.

The port array normally has periodic radial symmetry about an axis called the burner-axis. As can be seen in Fig. 7, for each case and after the potential core, the flows are dragged together along a "virtual centreline" of the numerical domain, specifically the burner-axis, creating a drag cone. When the ambient fluid is at rest or uniformly moving parallel to the burner-axis, the jets will be dragged into the axis combining their self almost as if their momentum flux fields were simply added or linearly overlapping and revelling the "drag cone". This cone is mainly produced by the mutual jets' interaction, however, this behaviour is reinforced and accentuated by the combustion, product of the reaction between air and fuel. Yimer et al. [13] call this structure the "flame necking cone" or FNC showing its characteristics until the air drag radius is practically imperceptible being conveniently determined through the using the surrounding gas minimum strain rate.[41–43]

Immediately after the potential core development, it is appropriate to assign a location range to the FNC where the mixing process is restricted to a distance where the air/fuel ratio is lean, but still the flame can take place. The result is, at some distance or height (Table 6: FL Heights and ranges) from the burner depending on the fuel flow, all traces of the original individual jets disappear and the assembly becomes a single round jet imperceptible from those individual nozzles jets from which it develops and comes directly. Because the supplied fuel-flow remains constant in each case, this structure becomes "self-preserved" until there is a variation of fuel concentration that does not allow to reach the lower flammability limit, LFL, or exceed the upper flammability limit, ULF. Now, the intermix jet behaves as if it came from a single point source or a single fuel nozzle and the individual details of the actual source, i.e. each nozzle may be negligible. However, the FNC seems to grow in a linear proportional way, which in summary indicates that the drag entrainment process inevitably occurs for each case [44]. Therefore, the major influences over the flame front will take place in this area. Table 6 presents dimensions of the characteristics of the zones.

Table 6: Dimensions of the characteristic of the flame zones L/d.

| Case | A | B | C | D | E | F |
|---|---|---|---|---|---|---|
| Fuel flow [cc/min] | 350 | 650 | 950 | 1200 | 1500 | 1800 |
| FNC Height | 2.05 | 2.89 | 3.57 | 3.94 | 4.66 | 5.22 |
| FL lower zone Height, $h_{FLl}$ | 1.15 | 1.11 | 1.05 | 1.05 | 1.03 | 1.03 |
| FL lower zone Range | $0<h_{FLl}>1.15$ | $0<h_{FLl}>1.1$ | $0<h_{FLl}>1.05$ | $0<h_{FLl}>1.05$ | $0<h_{FLl}>1.03$ | $0<h_{FLl}>1.03$ |
| FL upper zone Height, $h_{FLu}$ | 3.7 | 5.75 | 7.6 | 8.91 | 10.17 | 11.14 |
| FL upper zone Range | $0.3<h_{FLu}>4$ | $0.25<h_{FLu}>6$ | $0.21<h_{FLu}>7.81$ | $0.23<h_{FLu}>9.14$ | $0.27<h_{FLu}>10.44$ | $0.36<h_{FLu}>11.5$ |

There is no single parameter that defines flammability, but one that is relevant to gaseous mixtures is the flammability limit, which provides and defines the range of fuel concentrations, usually in percentage volume at 298 K, for flame ignite, burn and propagation occurs within a possible explosive reaction of a gaseous mixture in air when an external ignition source such as a spark is introduced. The LFL is the minimum limit of a combustible substance composition or concentration above which a flame is capable of propagating through a homogeneous mixture with air. Below the LFL, there is not enough fuel to cause ignition. The UFL is defined as the maximum one. With fuel concentration greater than the UFL, there is insufficient oxygen for the fuel to react and propagate away from the source of ignition until it is mixed with more oxygen.

The flammability limits, FLs, of chemical substances depend on many factors, including: flame propagation direction, mixture temperature and pressure, presence of fuel and oxidant concentrations flash point and $T_{ai}$ to mention few within safety specifications. Usually, the limits are experimentally obtained by determining the limiting mixture compositions between flammable and non-flammable mixtures [45] with the empirical representation with the eq. (15) and (16);

$$LFL_{T,P} = 1/2(C_{g,n} + C_{l,f}) \qquad (15)$$

$$UFL_{T,P} = 1/2(C_{g,f} + C_{l,n}) \qquad (16)$$

where the subscripts T and P indicate that is a function of temperature and pressure, respectively, commonly stated at 293K and 1atm; $C_{g,n}$, $C_{l,n}$ are the greatest and the least fuel concentration in an oxidant that is non-flammable; $C_{l,f}$, $C_{g,f}$ for flammable. The above criterion for flammability limit estimation is flame propagation from the point of ignition to a certain distance. The best-known experimental method using visual identification for measuring FLs of pure gases, pre-mixed air/gas as well as some gas mixtures, is that developed by the Bureau of Mines [46]. For this particular fuel mixture the LFL≈1.98 and UFL≈9.04 were computed form $C_{g,n}$, $C_{l,n}$, $C_{l,f}$, and $C_{g,f}$ values taken from those found in [46], respectively.

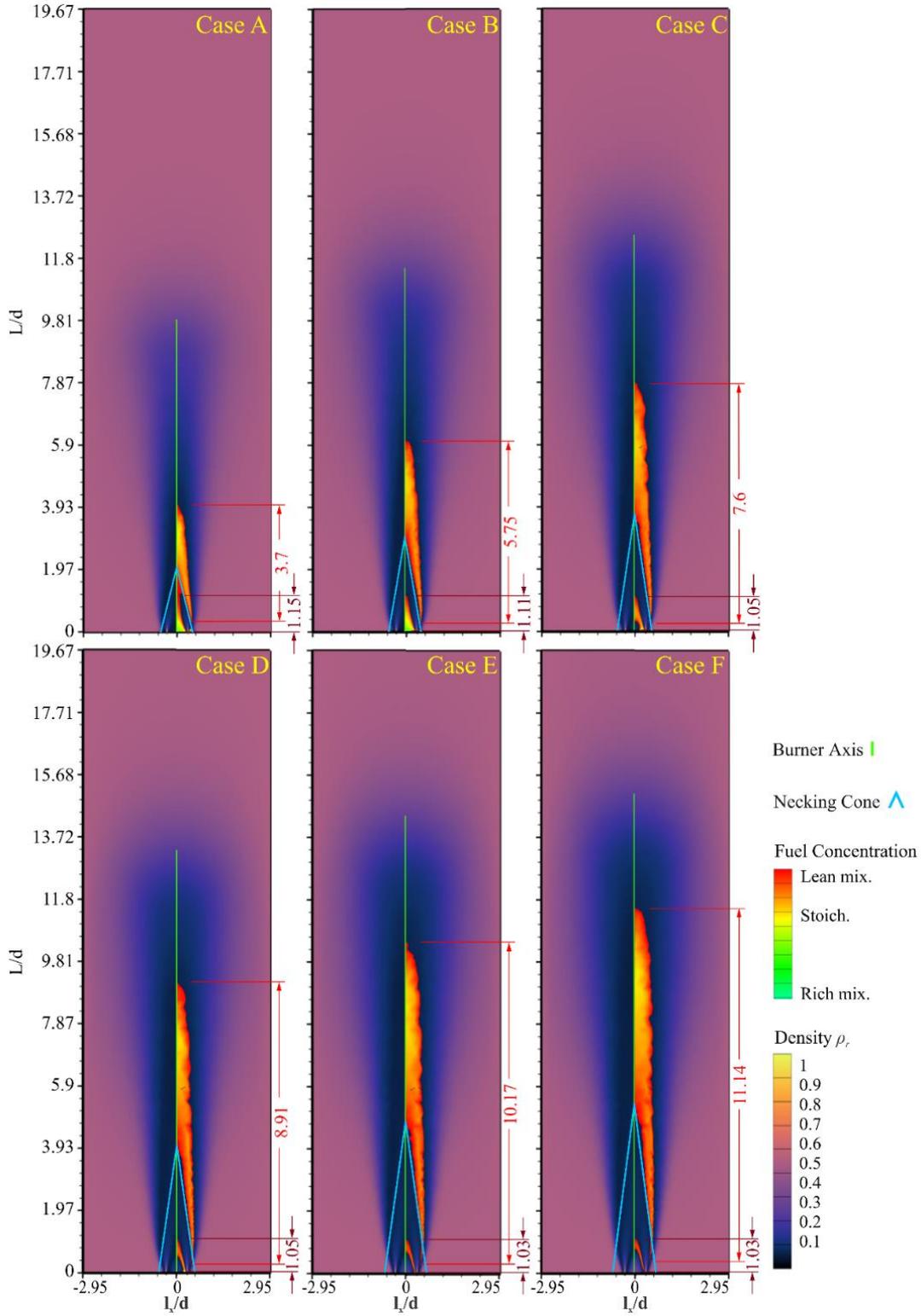

Fig. 7: Flame characteristics and measurements

The Fig. 7 illustrates the single nozzle behaviour, the burner-axis, the FNC and the FLs, overlapped on the reduced density profile in a cross-section on the central-plane along the Z-axis. The reduce density is $\rho_r = 2(\rho_f - \rho_a)/\rho_a$ where the sub-index $f$ and $a$ are for fuel and air respectively, where $\rho_f = 1.55 \text{kg/m}^3$ and

$\rho_a$=1.017kg/m$^3$. Since $\rho_r$ is based on a proportional mass value and for the main purpose of not being confused with FLs, this is considered to be the normalized density at which the fuel and air densities must be mass mixing ratios. It takes away the proportion of air/fuel mass already mixed, and the residue must be two times greater to be able to mix correctly with the surrounding air supplied. This is a rather simple and practical parameterization.

It is appreciable in Fig. 7 that, flame's FL length grow on the burner-axis as the mass flow of each case increases. This is because somehow the amount of surrounding air is reacting with the fuel somewhere in the lower part of the flame mixing zone causing the stoichiometric amount of fuel on the flame front not be linearly proportional to the range calculated. This behaviour can have several consequences, the principal one is that the flame front length, as well as the visible height and the flame luminous height, are directly harmed in the same proportion in which the FLs are reached.

For a single nozzle flame, the outside-air mixes from the periphery into the fuel jet. After the potential core and as the jet develops, the fuel concentration reaches the ULF, while, if there is a certain amount of fuel that does not react in the lower area of the flame, the fuel traces that are still dispersing can reach the necessary concentration of the lower limit as shown in the research by Kang et al. [47]. As a consequence, for a single nozzle configuration, the FLs are reached from inside the flame perpendicular to the fuel jet outlet direction, linearly and proportionally, when the mass flow increases, but not for the multi-nozzles-burners' configuration.

## 4.2   OUTER BOUNDARY GAS BEHAVIOUR

As mentioned in the previous afforded section, the most relevant zone in the mixing process of diffusion flames for this four-port multi-nozzle burner radial array is the FNC.

The Strain rate, $\sigma_K$, formulated by Seshadri and Williams [48], treats the mixing layer as a thin sheet within the mixing zone being a function of nozzle exit bulk flow velocities, distance-S, $\rho_f$ and $\rho_a$. With this formulation it is possible to determine part of the gases behaviour in the lower zone of the flame. Fig. 8 shows the $\sigma_K$, at 0L/d position in an XY plane cross-section of the outer boundary gas behaviour. In this plane are shown the streamlines of the surrounding air that is drawn from the burner edge into the cone. If the $\sigma_K$ remains within at an approximate range $0<\sigma_K>100s^{-1}$ the drag is moderate, allowing smooth air/fuel mixing. As this $\sigma_K$ increases, the surrounding air is forced to interact with the overall behaviour of the jets-array. Thus, the combustion process is carried out from the periphery towards the flame interior agreeing with the general unification behaviour of the multi-nozzle burner jets in which all traces of the original individual injectors disappear becoming a single round jet imperceptible from those individual nozzles' jets giving rise to a single flame front.

The FNC drag radius is approximately 20.5% greater than the gas burner radius for each case which radial distance is $r_d$=0.492l$_y$/d. This means that the flame development is similar for all cases because they have an analogous mixing process at least, within the range L/d<3.93. However, within the $r_d$, at which the nozzles are arranged and as the fuel flow increases, there is a $\sigma_K$ decrease over the XZ 45°angled-plane for each case. The comparison of the $\sigma_K$ values is shown in Table 7. This is because, when the planes are changed, the $\sigma_K$ is not influenced by the fuel jets' development that blends with the entrained air, not as it occurs in the XZ central-plane when both fluids "collide" and increase these values.

Table 7: Values of the $\sigma_K$ on the XY plane at 0L/d.

| Case | A | B | C | D | E | F |
|---|---|---|---|---|---|---|
| Fuel flow [cc/min] | 350 | 650 | 950 | 1200 | 1500 | 1800 |
| $\sigma_K$ at central-plane | 414.55 | 720.65 | 989.31 | 1116.55 | 1416.23 | 1596.88 |
| $\sigma_K$ at 45°angled-plane | 573.5 | 441.2 | 344.65 | 272.57 | 206.6 | 171.01 |
| FNC Range | \multicolumn{6}{c}{-0.706<l$_x$/d>0.706 and -0.706<l$_y$/d>0.706} |
| FNC Diameter | \multicolumn{6}{c}{1.41} |

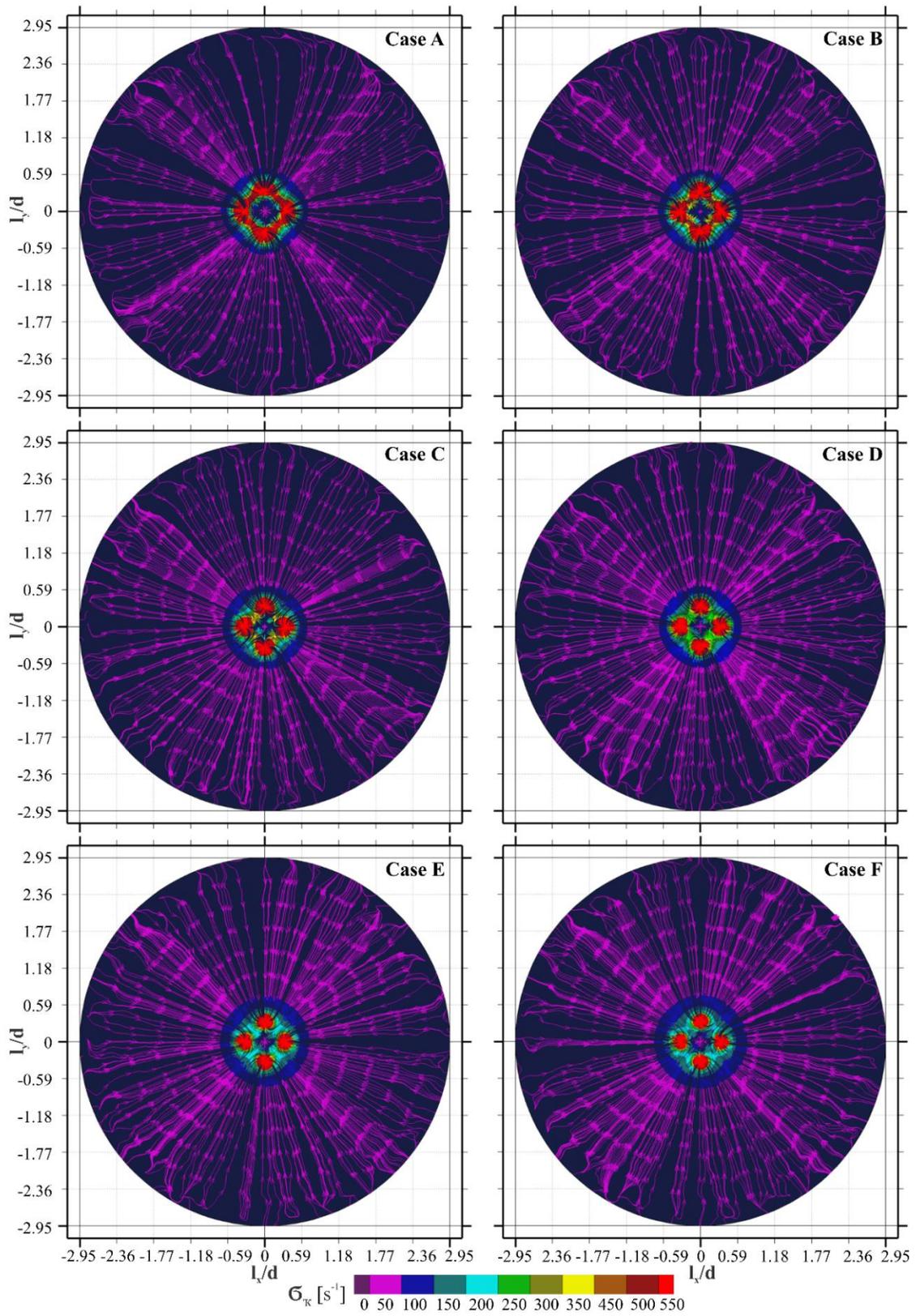

Fig. 8: $\sigma_\kappa$ at 0L/d position in a XY plane cross-section of the outer boundary gas behaviour.

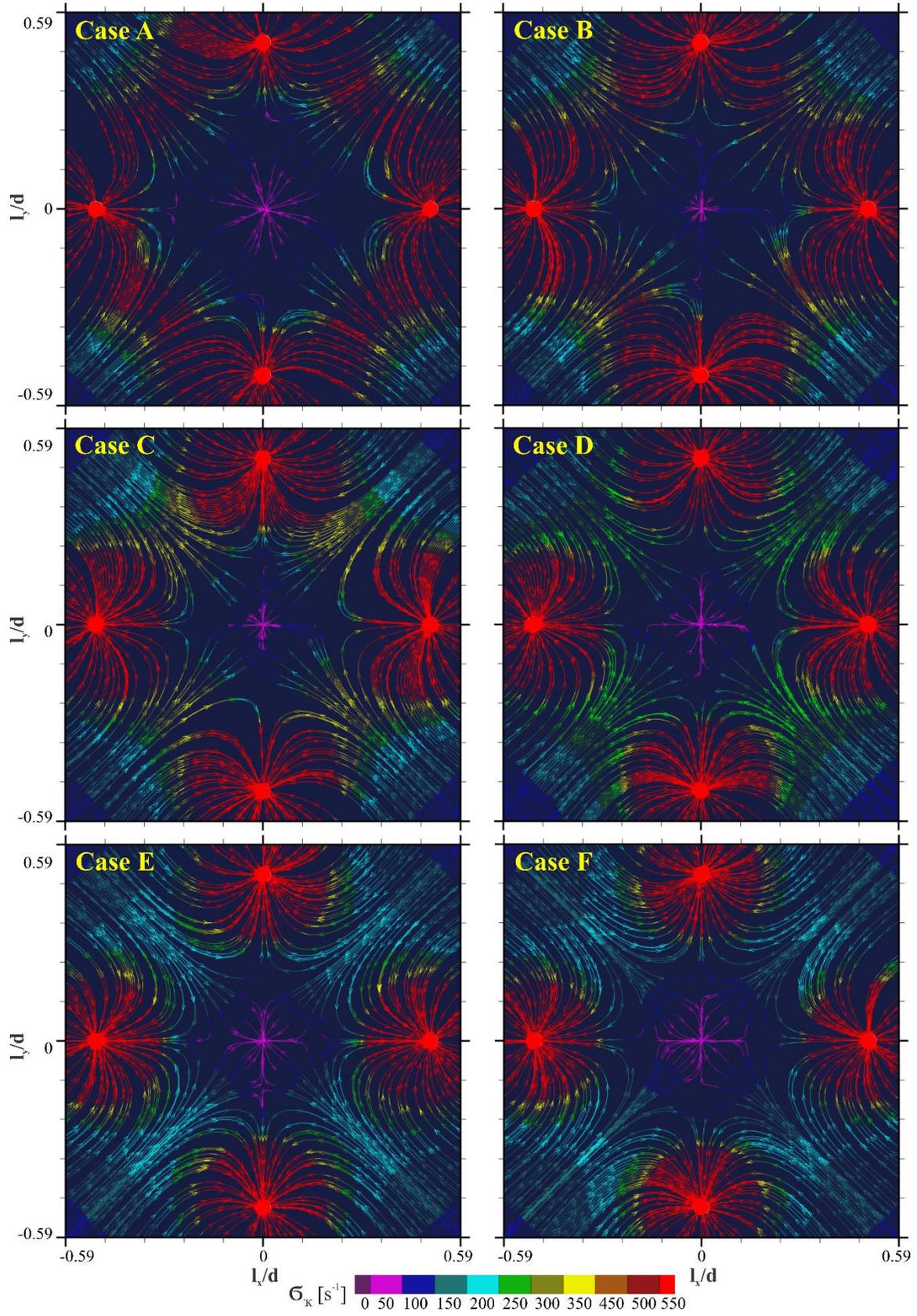

Fig. 9: Close-up at 0L/d position in a XY plane cross-section of the outer boundary gas behaviour.

Fig 9 exhibits the $\sigma_K$ close-up visualization at 0L/d position in a cross-section XY plane of the outer boundary gas behaviour. This figure confirms the evolution of the $\sigma_K$ values in the 0L/d position. Although the behaviour and mixture development in this position are explained, it does not conclusively explain the subsequent flame front development. That is, at different L/d positions over the XZ central-plane along the Z-axis, there is a $\sigma_K$ values increment in each case, but a decrement over XZ 45°angled-plane that could be contradictory.

The discrepancy between the $\sigma_K$ values at the different positions is mainly because there is a certain amount of air entrained from the periphery towards the fuel nozzle that filters between the distance-S, causing a mixture of air/fuel inside the FNC on the burner-axis along the Z-axis. The distance-S allows the air surrounding the jets, and later surrounding the flame front, to blend with the fuel, which are traces rich enough in oxygen to reach LFL. This area that develops on the burner-axis is known as the mixing length.

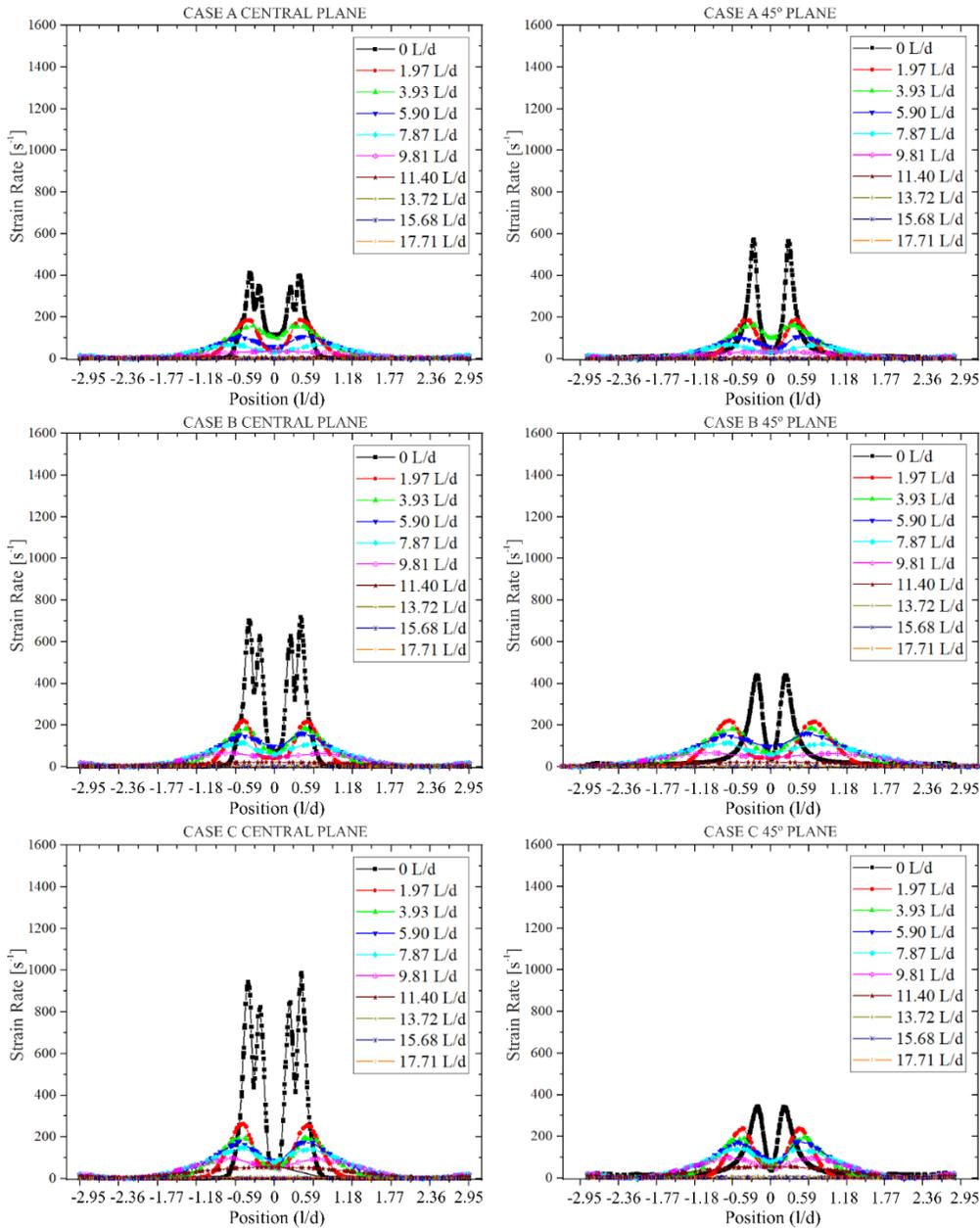

Fig. 10: Planes comparison of the $\sigma_K$ values at different position for case A to C.

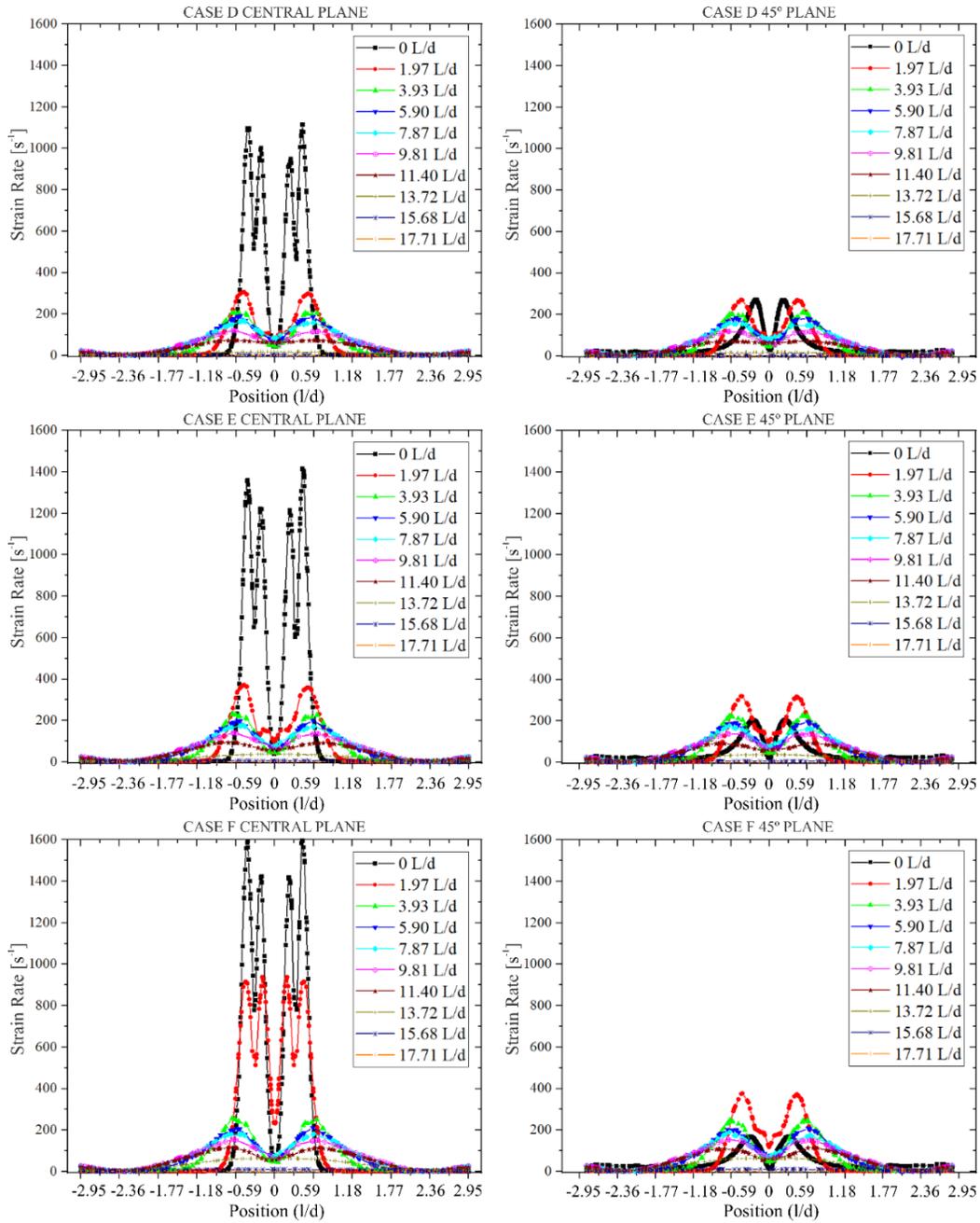

Fig. 11: Planes comparison of the $\sigma_\kappa$ values at different position for case D to F

The overall $\sigma_\kappa$ at different L/d positions are shown in Fig 10 and 11 showing that the mixing length occurs when $\sigma_\kappa<350s^{-1}$ according to the flame extinction Strain-rate [41–43]. In each case, the mixing length occurs up to L/d<3.93. If the $\sigma_\kappa<350s^{-1}$ at a position L/d<3.93, the surrounding air is still dragged in by peripheral nozzles jets. On the other hand, if the $\sigma_\kappa>350s^{-1}$ but at a position L/d>9.81, it is related to the combustion products dispersion. Similarly, if the $\sigma_\kappa<350s^{-1}$ at a position L/d>9.81, it means that there is no more air/fuel mixing process, but the air dragging effect from the flame edge continues, either driven by the flame front evolution or, driven by the combustion products dispersion. The resulting values are presented in Table 7.

Table 7: Position of the mixing length and combustion products dispersion L/d.

| Property | Location | A | B | C | D | E | F |
|---|---|---|---|---|---|---|---|
| *Mixing Length* | Central-plane | 1.97 | 1.97 | 1.97 | 3.93 | 3.93 | 3.93 |
| | 45°angled-plane | 1.97 | 1.97 | 1.97 | 1.97 | 1.97 | 3.93 |
| *Combustion products dispersion* | Central-plane | 9.81 | 9.81 | 11.4 | 11.4 | 13.72 | 13.72 |
| | 45°angled-plane | 9.81 | 9.81 | 11.4 | 11.4 | 13.72 | 13.72 |

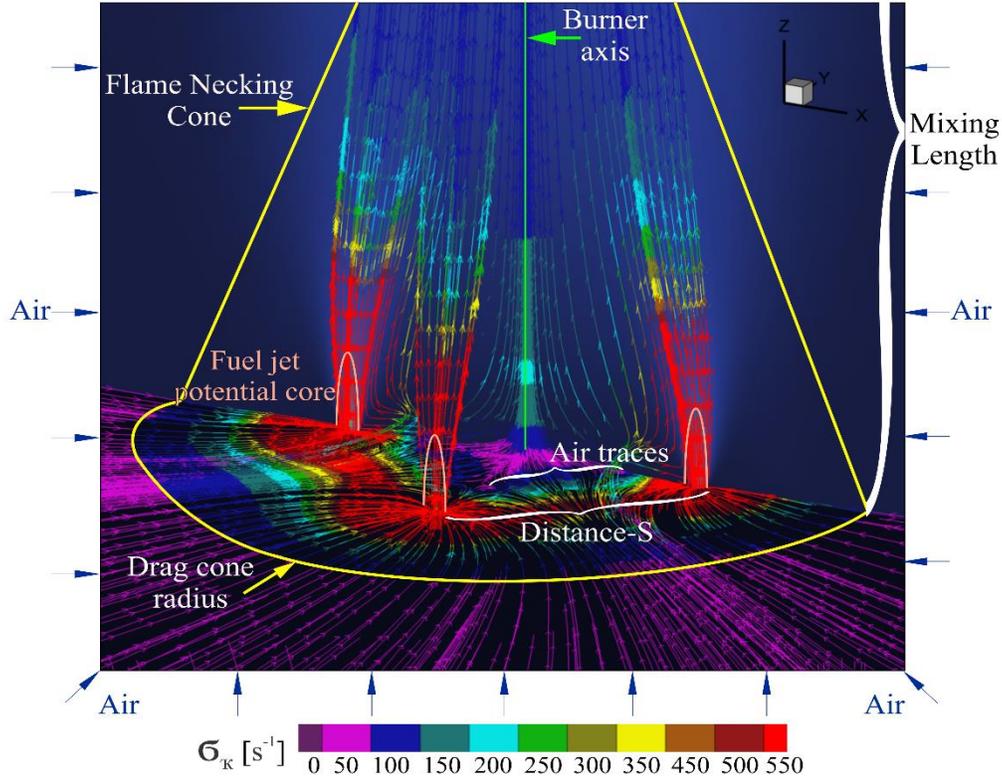

Fig. 12: Overall graphic summary of the detailed physical behaviour by zones.

Figure 12 shows a visual summary of the mixing process by zones and explained as follows: The fuel flow is introduced by the radially arranged nozzles with a distance-S between them. As the fuel is introduced, the characteristic that controls its diffusion is the potential core that each jet develops after the nozzles. Posterior the nozzles' exit, the momentum of each jet is enough to drag the air from the environment located at nozzles' outlet. When the jets are diffused, this movement drags the air around the jets and the periphery. This mechanism causes the fuel and air to coincide in the mixing zone, developing its characteristic length. This mixing occurs similarly for each jet. However, in the space between them, distance-S, a certain amount of air is filtered. The momentum of these air traces is sufficient to reach the burner-axis without mixing with the diffused fuel that is also towards the burner-axis due to self-preserving drag. Due to the mixing process is driven by this drag, the air/fuel mixture is smooth, achieving flammability limits at a lower location or very close to the burner surface.

The entire mixing-mechanism driven by the jets culminates with their self-interaction, developing the flame necking cone. Any concentration of air that falls into this flame cone will be entrained, contributing both to the jet's development and the overall flame's structure. The reach of cone radius is limited only by the radial distance of the jet array and the nozzles distance-S. Its height will depend solely by the amount of fluid mass flow that exist from the nozzles. All this behaviour better describes and complements the physical phenomena, why as the nozzles' radial-distance increases, the characteristic lengths, in general, decrease as described Lenze et al and Kim & Lee [18,49]. It also explains more precisely why, by extending or diminishing the distance-S that is achieved by increasing the number of nozzles or the separation between them respectively for to maintain the interaction between the jets it is necessary to step up the amount of fuel coming out from nozzles causing the lift-off to exist as explained in some studies [50,51]. Therefore, there is a well-defined fuel mass flow range for each distance and the fuel-nozzles number. For the present study distance-S and 4 nozzles are within the range of 950cc/min to 1200cc/min, which are case C and D respectively. And finally, this information can be used to optimize and improve the efficiency of diffusion flames in the combustion processes of current multi-port burner technologies.

## 5    CONCLUSIONS

This work shows the combustion mixing process for a 4-port array for six different fuel flows which have been numerically predicted using a commercial CFD code and compared against experimental measurements with a statistically significant adjustment based on the magnitude of temperature.

The flow from a multi-port burner becomes a single self-conserving round jet due to the development of the flame necking cone whose drag radius is approximately 20.5% greater than the gas burner radius for each case which is $r_d=0.492l_y/d$. This means that the flame development is similar for all cases because they have an analogous mixing process, within the range $L/d<3.93$.

The $\sigma_\kappa$ values reveal that at the different positions air-traces are rich enough in oxygen to reach LFL. These traces are entrained from the periphery towards the fuel nozzle that filters between the distance-S, causing a mixture of air/fuel inside the FNC. The mixing length occurs when $\sigma_\kappa<350s^{-1}$. In each case, the mixing length occurs up to $L/d<3.93$. If the $\sigma_\kappa<350s^{-1}$ at a position $L/d<3.93$, the surrounding air is still dragged in by peripheral nozzles jets. On the other hand, if the $\sigma_\kappa>350s^{-1}$ but at a position $L/d>9.81$, it is related to the combustion products dispersion. Similarly, if the $\sigma_\kappa<350s^{-1}$ at a position $L/d>9.81$, there is no more air/fuel mixing process, but the air dragging effect from the flame edge continues, either driven by the evolution flame front or, driven by the combustion products dispersion. Therefore, for the present study, the ideally stable mixing process fuel flow is within the range of 950 cc/min to 1200 cc/min corresponding to case C and D, respectively. And finally, this information can be used to optimize and improve the efficiency of diffusion flames in the combustion processes of current multi-port burner technologies.

## 6    ACKNOWLEDGEMENTS

The authors appreciate the support provided by the National Autonomous University of Mexico and grants provided by the National Council of Science and Technology of Mexico. In addition, we appreciate the support provided by Laboratory of Applied Thermal and Hydraulic Engineering, Superior School of Mechanical and Electrical Engineering of the National Polytechnic Institute. The authors are grateful to the support, contributions made and the information provided by the National Institute of Nuclear Research.

## 7    FUNDING

The authors appreciate the support provided by the National Autonomous University of Mexico Engineering Institute through the project DGAPA-PAPIIT IN112419. The research work described in this paper was supported by the grants from the National Council of Science and Technology of Mexico (CONACyT) and the Mexican Ministry of Energy (SENER), as well as the resources provided by the Applied Thermal and Hydraulic Engineering Laboratory of the National Polytechnic Institute of Mexico.

## 8 AUTHOR CONTRIBUITIONS

**M. De la Cruz-Ávila:** Conceptualization, Investigation, Methodology, Software, Validation, Formal Analysis, Visualization, Resources, Writing – Original Draft, Review & Editing, Project Administration, Funding Acquisition;
**J.E. De León-Ruiz:** Conceptualization, Methodology, Validation, Formal Analysis, Investigation, Data Curation, Visualization, Writing – Original Draft, Review & Editing, Resources;
**E. Martinez-Espinosa:** Writing – Review & Editing, Project Administration;
**I. Carvajal-Mariscal**: Methodology, Validation, Formal Analysis, Investigation, Resources, Writing – Review & Editing, Supervision, Project Administration;
**L. Di G. Sigalotti:** Validation, Writing – Review & Editing.